# Moiré excitons in generalized Wigner crystals

Jing-Yang You[1,†], Chih-En Hsu[1,2,†], Zien Zhu[1], Benran Zhang[1,3], Ziliang Ye[4,5], Mit H. Naik[6], Ting Cao[7], Hung-Chung Hsueh[2], Steven G. Louie[8,9,*], Mauro Del Ben[10,*], and Zhenglu Li[1,*]

[1]Mork Family Department of Chemical Engineering and Materials Science, University of Southern California, Los Angeles, CA, USA

[2]Department of Physics, Tamkang University, Tamsui, New Taipei, Taiwan

[3]Thomas Lord Department of Computer Science, University of Southern California, Los Angeles, CA, USA

[4]Department of Physics and Astronomy, The University of British Columbia, Vancouver, BC, Canada

[5]Quantum Matter Institute, The University of British Columbia, Vancouver, BC, Canada

[6]Department of Physics, The University of Texas at Austin, Austin, TX, USA

[7]Department of Material Science and Engineering, University of Washington, Seattle, WA, USA

[8]Department of Physics, University of California at Berkeley, Berkeley, CA, USA

[9]Materials Sciences Division, Lawrence Berkeley National Laboratory, Berkeley, CA, USA

[10]Applied Mathematics and Computational Research Division, Lawrence Berkeley National Laboratory, Berkeley, CA, USA

[†]These authors contributed equally.

[*]Email: sglouie@berkeley.edu (S.G.L.), mdelben@lbl.gov (M.D.B.), zhenglul@usc.edu (Z.L.)

**Abstract**

Moiré superlattices of transition-metal dichalcogenide bilayers host strong Coulomb interactions residing in narrow electron bands, leading to correlated insulating states at fractional carrier doping densities, known as generalized Wigner crystals. In excited states, the formation of moiré excitons is expected to be fundamentally shaped by the Wigner-crystal ground states, manifesting an intricate interplay between electronic and excitonic correlations. However, the microscopic description of these Wigner crystalline excitons (WCEs) remains elusive, largely subject to speculations, and is further needed for the understanding of exotic excitonic phases (e.g., exciton insulators and exciton density waves) and their unique properties (e.g., anomalous exciton diffusion). Here, using first-principles many-body $GW$-Bethe-Salpeter-equation calculations, we directly reveal the internal structures of WCEs in angle-aligned $MoSe_2/MoS_2$ moiré heterostructure at hole fillings of 1/3 and 2/3. Our results unveil the propagation of correlation effects from the ground state to excited states, shaping the real-space characteristics of WCEs. The strong two-particle excitonic correlations dominate over the kinetic energy of free electron-hole pairs, in analog to the strong single-particle correlations of flat bands. We propose that such unusual excited-state correlation effects of WCEs can be experimentally probed by photocurrent tunneling microscopy. Our work provides a microscopic understanding of strongly correlated WCEs, suggesting them as a highly tunable mixed boson-fermion platform to study many-body interactions and phenomena.

**Main**

Generalized Wigner crystals[1-3] emerge in moiré superlattices of transition-metal dichalcogenide (TMD) bilayer systems upon carrier doping due to strong electron correlation effects in flat bands, where the kinetic energy of the carriers is suppressed by the modulation of moiré potentials[4-21]. The Wigner-crystal electronic states host exotic quantum phenomena[22-27] by forming new charge orders at fractional carrier fillings (e.g., 1/3, 2/3, 2/5, etc.), breaking the translational symmetry of the moiré unit cells. Meanwhile, in atomically thin two-dimensional (2D) materials, the reduced dielectric screening further enhances the Coulomb interaction between excited electrons and holes, leading to the formation of strongly bound excitons[8,28]. In this work, we investigate the intriguing properties of Wigner crystalline excitons (WCEs) – excitons arising from correlated Wigner crystal states – where moiré superlattice potential, Wigner crystallization, and strong excitonic effects coexist and jointly shape the many-body interaction landscape. Correlated insulating exciton phases[9,14-18] and Hubbard-type exciton-exciton interactions[16,19] have been observed, which are strongly modulated by the carrier-doping-induced Wigner-crystal or Mott-insulating states. Moreover, WCEs show unusual diffusive and dynamical behaviors[19-21]. The microscopic nature of these strongly correlated excitonic states dictates the formation of exotic orders (e.g., via excitonic dipolar interactions[16,29,30]), but its understanding remains largely elusive. Current interpretation of experiments often relies on qualitative speculations based on assumptions. In the case of moiré excitons arising from a conventional semiconducting ground state (i.e., not in a doping-induced correlated insulating state), the excitonic internal structure can be understood as inheriting those from free electron-hole pairs (i.e., characters of the valence and conduction bands) shaped by moiré superlattice potentials. However, it is questionable whether the same picture applies to WCEs and what role does the strong Wigner-crystal correlation play.

Addressing the nature and internal structure of WCEs thus becomes an important and timely quest, calling for microscopic first-principles-based studies. *Ab initio GW*-Bethe-Salpeter-equation (*GW*-BSE) calculations[31-36] provide accurate descriptions of excitons and their properties, but face formidable computational challenges for large-scale moiré systems (typically, with a few thousand atoms per unit cell). To enable such *GW*-BSE calculations, previous studies[8,37,38] have successfully developed a uniform (non-moiré) dielectric screening approximation for excitons in undoped moiré superlattices, and predicted novel charge-transfer moiré excitons[8] that were later confirmed and visualized experimentally using a photocurrent tunneling microscopy[38] (PTM) setup. However, the formation of WCEs is rooted from carrier-doped correlated insulating ground states, where even very low carrier concentrations can cause drastic changes in the electronic and excitonic properties of 2D materials[39-41]. Moreover, the carrier doping is spatially nonuniform due to the formation of Wigner-crystal states. Here, we have enabled direct *GW*-BSE calculations of WCEs in an angle-aligned $MoSe_2/MoS_2$ moiré heterostructure at hole fillings of $v_h$ = 1/3 and 2/3, by incorporating a detailed inhomogeneously doped dielectric environment of the generalized Wigner crystals. The computation of WCEs requires enlarged supercells of the Wigner crystal composing multiple original moiré unit cells, pushing the system size to ~10,000 atoms. The computational bottlenecks from both the unavoidable nonuniform dielectric environment and the increased system size are efficiently addressed in this work through a series of methodological and algorithmic developments.

Our results have unambiguously revealed the internal structure of WCEs. We discover that the ground-state correlation effects shaping the Wigner crystals propagate into the excited states, dictating the formation of unique characteristics of WCEs. In the fractionally hole-doped $MoSe_2/MoS_2$ moiré superlattices, the position of the excited electrons closely follows the real-space distribution of the excited holes, which are pinned by the Wigner-crystal ground state, avoiding the position of the doped holes in the valence bands. Namely, the excited electrons are strongly correlated with the excited holes, instead of being shaped by conduction-band wavefunctions as in the prevailing picture of conventional moiré excitons. The electron-hole interaction is found to be much stronger than the kinetic energy of free electron-hole pairs (i.e., the interband transition energy dispersion) by over an order of magnitude. In analog to the ground-state Wigner crystals, the formation of excited-state WCEs in this system is dominantly correlation driven. We further propose a feasible PTM experiment[38] to probe such strongly correlated nature of WCEs.

The angle-aligned (with zero-degree twist angle) $MoSe_2/MoS_2$ moiré superlattice is constructed in our calculations (Fig. 1a) with a moiré unit cell consisting a layer of 23×23 $MoSe_2$ supercell and a layer of 24×24 $MoS_2$ supercell to accommodate the lattice mismatch between the two layers (see Supplementary Information). As expected, the system undergoes a structural reconstruction due to the stacking energy differences[42] in various regions. We first investigate the single-particle electronic structure using density functional theory[43] (DFT) and $GW$ method[35]. The band gap (including the $GW$ correction to DFT) of the heterostructure shows a type-II band alignment, where the valence band top (VBT) is mostly from the Γ-valley of the $MoSe_2$ layer (with some hybridization with the $MoS_2$ layer), and the conduction band bottom (CBB) is mostly from the K-valley of the $MoS_2$ layer. For the undoped $MoSe_2/MoS_2$ superlattice, the band structure in the moiré unit-cell Brillouin zone (Fig. 1b) shows a singly degenerate band as VBT, and two nearly degenerate bands as CBB. These bands are isolated and quite flat, with a bandwidth of ~1 – 2 meV, suggesting conditions for the emergence of strongly correlated phases if doped with carriers. Introducing carrier doping into the flat bands causes strongly correlated insulating charge-ordered states such as the generalized Wigner crystals and Mott insulators, which have been consistently observed experimentally. Here, the intersite Coulomb repulsion between different stacking sites suppresses the kinetic energy of the doped carriers and isolates them spatially to reduce the total energy of the system[1-3]. In this work, we mimic the ground-state correlation physics and the electronic structure using the DFT+$U$ method[44,45] with imposed real-space symmetry breaking (see Supplementary Information), which allows us to create generalized Wigner-crystal ground states as well as their single-particle excitations for further computing of properties of WCEs using the $GW$-BSE approach. Note that the ground-state wavefunction of generalized Wigner crystals can be reasonably represented by a single Slater determinant[46] (e.g., using the Hartree-Fock method[46-48]), justifying the applicability of DFT+$U$ and $GW$-BSE approaches in this work.

Fractional carrier doping per moiré unit cell is achieved by the construction of enlarged new supercells on top of the neutral moiré crystal structure. For $v_h$ = 1/3 and 2/3 in the $MoSe_2/MoS_2$ heterobilayer, we adopt a $\sqrt{3} \times \sqrt{3}$ supercell of moiré unit cells (Fig. 1a), consisting of 9,945 atoms. In the new supercell Brillouin zone, the pristine moiré band corresponding to VBT is folded into a triplet and the two bands of the CBB complex is folded into a sextuplet. Figs. 1c and 1d show the density maps of the wavefunctions of the CBB and VBT of the undoped superlattice, respectively, where the wavefunctions are mostly located at $B^{Se/Mo}$ stacking sites, displaying the translational periodicity of the undoped moiré unit cells. To induce symmetry breaking and simulate the effect of Coulomb interaction that leads to localization and correlation for the doped carriers, we add an onsite $U$ to the $d$ orbitals of the Mo atoms in the $MoSe_2$ layer around the $B^{Se/Mo}$ stacking sites selectively (see Supplementary Information). Upon $v_h$ = 1/3 hole doping (per pristine moiré unit cell), one band in the triplet manifold is emptied out, leaving an unoccupied singlet band and occupied doublet bands in the VBT complex, opening a mini band gap of 4.7 meV (Fig. 1e), consistent with the experimental estimation of ~0.5 – 50 meV in different systems[22,27,49,50]. The corresponding wavefunction of the occupied doublet bands shows a symmetry-broken charge order of a hexagonal lattice, resembling the Wigner-crystal states with one missing electron pocket per three electron pockets of VBT. Similarly, with $v_h$ = 2/3, the VBT is split into unoccupied doublet bands and an occupied singlet band by a gap of 4.8 meV, where the occupied singlet band shows a triangular lattice of Wigner-crystal charge order with two out of three electron pockets missing (Fig. 1f).

The existence of the correlated insulating ground states at specific fractional doping has fundamental implications for the excited states, relevant to a broad range of phenomena. From a simple band-to-band transitions perspective as depicted in Fig. 1**g**, some interband transitions become forbidden due to Pauli blocking. Further, considering the symmetry-breaking nature of these ground states, the formation of WCEs becomes rather intriguing and is the focus of this work. In general, the quasiparticle states outside of the VBT complex (including those in CBB) can sense the correlated charge orders within the VBT complex upon hole doping, with their energy shifts expected to be smaller than the correlation gap. However, because the occupations of the non-VBT states remain unchanged, their small modifications from the doping effects are expected to be negligible in the exciton formation processes, as compared to those from the VBT complex where a drastic occupation change takes place upon doping, both in real space and energy space. Thus, to focus on the formation of WCEs, in our construction of the electronic structure of generalized Wigner crystals, we restrict the effects of correlation only within the VBT complex (see Supplementary Information).

With a given electronic structure, solving for WCEs for large moiré structures using the first-principles *GW*-BSE approach is further a considerable computational challenge. Prior studies[8,37,38] of moiré excitons in TMD bilayers, with conventional semiconducting (undoped) ground states, required the development of a pristine unit-cell matrix projection (PUMP) method[8], which significantly reduces the problem size. In the PUMP method[8], the dielectric matrix of the moiré superlattice is approximated by that of the pristine unit cell, which allows the electron-hole interaction to be expanded using non-moiré Bloch states. The uniform dielectric screening approximation is valid and excellent for moiré excitons[8] in undoped systems with a conventional ground state. However, the WCEs of interest here arise from doped ground states with the doped carriers distributed inhomogeneously and manifested as generalized Wigner crystals. It is found that the excitonic properties[40,41], as well as the quasiparticle band gaps[39], are quite sensitive to carrier doping in 2D semiconductors, albeit the exciton excitation energies stay less sensitive due to a cancellation effect of the reduction in both band gaps and exciton binding energies[40,41].

In this work, we aim at capturing the sensitive and subtle correlated doping effects (beyond the realm of PUMP) that are central to the WCEs. We have addressed these computational challenges, enabling direct *GW*-BSE calculations. In particular, we capture the screening effects induced by the Wigner-crystal charge orders and the moiré superlattice potentials by including the dominant and detail-sensitive low-energy transitions explicitly in the polarizability, but with the less-dominant background high-energy transitions constructed from pristine unit cell calculations (see Supplementary Information). We then compute the electron-hole interaction kernel and solve the BSE (the exciton Hamiltonian) directly in the Hilbert space of the $\sqrt{3} \times \sqrt{3}$ moiré supercell with no more additional approximations beyond the standard formalism[31-34]. Through redesign of code implementations and adaption to graphic-processing-unit accelerators[51], direct *GW*-BSE calculations of ~10,000 atoms become feasible on exascale supercomputing facilities.

Figs. 2**a** – 2**c** show the calculated excitation energies of the moiré excitons. As a function of doping (i.e., $v_h$ = 0, 1/3, 2/3), clear reductions in both the quasiparticle band gaps and exciton binding energies happen simultaneously, due to the enhanced screening from the doped carriers. Consequently, the excitation energies (~1.2 eV) of the lowest-energy excitons (composing of interband transitions across the main gap) are much less sensitive to doping, due to the cancellation effects reported earlier[40,41] (see Supplementary Information). Our calculations show that the lowest-energy excitons have negligible oscillator strengths and therefore are optically dark. This is because the main compositions of the VBT and CBB in the supercell Brillouin zone are derived from different parts of the Brillouin zone of the pristine unit cells (the Γ-valley and K-valley, respectively), inheriting the momentum-forbidden nature of the optical transitions. However, the dark nature of WCEs is essential for their observation in PTM experiments, which require long-lived (dark) low-energy excitons[38], as discussed later.

The real-space behavior of the wavefunction of an exciton state *S* depends on both the coordinates of the excited hole ($\mathbf{r}_h$) and the excited electron ($\mathbf{r}_e$), and can be expanded in the basis of free electron-hole pairs as $\chi^S(\mathbf{r}_e, \mathbf{r}_h) = \sum_{vc\mathbf{k}} A_{vc\mathbf{k}}^S \psi_{c\mathbf{k}}(\mathbf{r}_e) \psi_{v\mathbf{k}}^*(\mathbf{r}_h)$, where $A_{vc\mathbf{k}}^S$ is called the **k**-space envelope function from the solution of the BSE, and $\psi_{c\mathbf{k}}(\mathbf{r})$ and $\psi_{v\mathbf{k}}(\mathbf{r})$ are the Bloch wavefunctions of the conduction (*c*) and valence (*v*) bands, respectively. To directly visualize the internal structure of WCEs, we plot the hole and electron densities, with their counterpart coordinate integrated out, i.e., $\rho_h^S(\mathbf{r}_h) = \int |\chi^S(\mathbf{r}_e, \mathbf{r}_h)|^2 d\mathbf{r}_e$ and $\rho_e^S(\mathbf{r}_e) = \int |\chi^S(\mathbf{r}_e, \mathbf{r}_h)|^2 d\mathbf{r}_e$, respectively. At different hole doping concentrations, the excited hole densities $\rho_h^S(\mathbf{r}_h)$ of the excitons (Figs. 2**d** – 2**f**) resemble closely the densities of the occupied VBT manifolds (Fig. 1**d** – 1**f**). This behavior is natural because only the occupied states in VBT are available to host excited holes of low-energy excitons. However, remarkably, the excited electron densities $\rho_e^S(\mathbf{r}_e)$ (Figs. 2**d** – 2**f**) also closely follow the hole densities, instead of the conduction band densities (Fig. 1**c**). Our direct *GW*-BSE calculations show that the excited electrons and holes at $v_h$ = 1/3 and 2/3 are completely correlated in WCEs, where the symmetry-breaking ground-state correlation effects of Wigner crystals (a charge ordered phase) propagate into the excited states.

In Figs. 2**g** – 2**i**, we show the decomposition of WCEs into electron-hole band pairs, $P_{vc}^S = \sum_{\mathbf{k}} |A_{vc\mathbf{k}}^S|^2$. To simplify discussions, the *c-v* band indexing is kept the same as in the undoped case. Indeed, as expected, the low-energy excitons are mostly composed of transitions between VBT and CBB, along with minor contributions from higher-energy electron-hole pairs. Here, our analysis mostly focuses on exciton states in the optical energy range (~1 eV). Upon hole doping, e.g., $v_h$ = 2/3, the formation of a correlated gap of 4.8 meV within the VBT manifold gives rise to additional very low-energy excitons, with excitation energy at 1.8 meV based on our *GW*-BSE calculations. However, this very low transition energy is drastically different from those of the exciton states arising from the main fundamental gap; hence essentially, these two classes of excitations do not mix with each other (see Supplementary Information).

The strongly correlated nature between the excited electron and hole densities of WCEs (with the excited electron density differing from the conduction band density) suggests that the electron-hole attractive interaction dominates over the kinetic energy of free electron-hole pairs in forming the excitonic states. To dig deeper into the driving factor of this behavior of the WCEs, we perform auxiliary calculations with a scaling factor $\alpha$ to the electron-hole interaction. Specifically, we construct a Bethe-Salpeter Hamiltonian of the form $\hat{H} = \hat{T} + \alpha \hat{K}$ with various values of $\alpha \in [0,1]$, where $\hat{T}$ is the kinetic energy of free electron-hole pairs and $\hat{K}$ is the full electron-hole interaction kernel. ($\alpha = 1$ corresponds to the real physical system.) Fig. 3**a** shows the resulting electron and hole densities of the WCEs with varying $\alpha$ for $v_h$ = 2/3. At $\alpha = 0$, there is no electron-hole interaction, and the excitations are just free electron-hole pairs. Accordingly, the excited hole and electron densities follow the charge densities of the band states of the VBT and CBB complexes, respectively. As $\alpha$ increases, the electron-hole interaction plays an increasingly important role in the formation of WCEs, and a broken translational symmetry starts to develop in $\rho_e^S(\mathbf{r}_e)$ (i.e., moving away from the crystalline translational symmetry of the undoped system) following the ground-state Wigner-crystal density distribution, even though the carrier density of CBB itself displays the translational symmetry of moiré unit cells. We emphasize that the nature of WCEs is thus intrinsically shaped by the electron-hole interaction, extending the implications of the strong correlation of Wigner crystals into the low-energy exciton states.

Fig. 3**b** shows the line profiles of $\rho_h^S(\mathbf{r}_h)$ and $\rho_e^S(\mathbf{r}_e)$ for various $\alpha$ values. The development of translational symmetry-breaking Wigner-crystal features in $\rho_e^S(\mathbf{r}_e)$ induced by electron-hole correlations is clearly seen. Moreover, Fig. 3**c** shows the integrated pocket density at the indicated regions I, II, and III. We note that even at $\alpha = 0.05$, 98% of the full WCE features in the excited electron distribution are recovered, compared with $\alpha = 1.0$ (the true electron-hole interaction strength), suggesting a dominant role of the electron-hole interaction. As shown in Figs. 2**g** – 2**i**, the main compositions of the lowest-energy WCEs are from the VBT and CBB complex, which have a kinetic energy of ~2–3 meV (bandwidth of the joint density of states). Because the kinetic energy term in the BSE is so small, the electron-hole interaction can be estimated to be on the order of exciton binding energy ($E_B$), that is $E_B$ = 118 meV and 98 meV for $v_h$ = 1/3 and 2/3, respectively. These analyses show that the electron-hole interaction is over one order of magnitude stronger than the kinetic energy, utterly dominating the formation of WCEs. Note that the conventional moiré excitons at zero doping ($v_h$ = 0) are also fundamentally correlation-driven with a large exciton binding energy of 419 meV (Fig. 2**a**) and a small kinetic energy mostly from VBT and CBB (Fig. 2**g**). However, both kinetic-energy-dominant and interaction-dominant pictures yield the same internal exciton structure (holes following VBT and electrons following CBB) for the undoped case in this MoSe$_2$/MoS$_2$ superlattice; hence the electron-hole correlation-dominant nature becomes unidentifiable from the internal structure of the exciton in this case. In contrast, for WCEs in the doped superlattices, the correlation-driven nature is clearly distinguishable, leading to unique and experimentally verifiable exciton internal structures.

Here, we propose an experimental setup to unambiguously demonstrate the strongly correlated nature of WCEs, based on the recently developed PTM technique[38] (Fig. 4**a**). In such PTM experiment, the moiré superlattice first needs to be electrostatically gated into a Wigner crystal via doping on one side of the gap. In this work, we consider hole doping into VBT. Then, a continuous pump laser with frequencies above the band gap could be applied to excite the system. The excitations would relax to the lowest-energy WCE states arising from the main gap, which are long lived due to their optically dark nature[38]. The participation of excitons from the correlation gap (with ~2 meV excitation energy) in the relaxation process should be minimal because of the large energy difference (>1 eV). Then a scanning tunneling probe can be used to draw hole and electron currents associated with excitons with the laser on (see Supplementary Information), and map out the real-space distribution of $\rho_h^S(\mathbf{r}_h)$ and $\rho_e^S(\mathbf{r}_e)$ through varying the tip bias[38]. The visualization of the excited-state charge densities is intrinsically enabled by the long lifetimes of dark excitons, where the excited electrons and holes can tunnel into the STM tip before the excitations decay. This principle has been successfully demonstrated in the observation of theoretically predicted dark low-energy charge-transfer moiré excitons in near-60° twisted WS$_2$ bilayers[38]. In Fig. 4**b**, we present a series of simulated PTM spectra to illustrate this behavior qualitatively for the MoSe$_2$/MoS$_2$ superlattice with varying tip bias in the vicinity of $V_0$, which corresponds to the voltage where the integrated electron and hole currents balance out, resulting in zero net photocurrent. The specific value $V_0$ depends on sample condition, tip configuration, work function difference between the tip and back gate, as well as the back gate voltage[38]. The varying range of the tip bias should be of the order of ~100 meV so that only a weak perturbation is exerted on the excitons[38]. The strong correlation nature of WCEs directly manifests as a continuous onsite change in the magnitude and sign of the tunneling current, meaning that the excited electrons are bound to the sites of excited holes. On the other hand, the uncorrelated free electron-hole density map can be constructed from standard scanning tunneling spectroscopy measurements (laser off) of the VBT and CBB states as a spectrum difference (Fig. 4**c**), which shows a strong contrast in the spatial distribution compared to that of the WCEs.

Our first-principles-based descriptions of the electronic structure and WCEs in generalized Wigner crystal phase of TMD moiré structures provide a solid foundation for the microscopic interpretation and clarification of a range of phenomena. Instead of a kinetic-energy-driven picture, because of the flat bands, WCEs emerge as strongly correlated electron-hole pairs, pinned by the ground-state generalized Wigner crystals. This study paves the way to investigate intriguing mixed boson-fermion systems and design new optoelectronic sensing devices and programmable quantum materials.

**Acknowledgements**

Z.L., M.D.B., and S.G.L. acknowledge primary support by the Center for Computational Study of Excited-State Phenomena in Energy Materials (C2SEPEM) at the Lawrence Berkeley National Laboratory (LBNL), which is funded by the U.S. Department of Energy (DOE), Office of Science, Basic Energy Sciences, Materials Sciences and Engineering Division under Contract No. DEAC02-05CH11231, as part of the Computational Materials Sciences Program, which provided the development of methods and advanced codes. Work at the University of Southern California (USC) (J.Y.Y., C.E.H., Z.Z., B.Z., Z.L.) is partly supported by the Zumberge Preliminary Studies Research Award from USC. C.E.H. and H.C.H. acknowledge the support by National Science and Technology Council (NSTC) under Grant No. 113-2112-M-032-013. Z.Y. is supported by the Natural Sciences and Engineering Research Council of Canada and the Canada Research Chairs Program. M.H.N. acknowledges the start-up support provided by The University of Texas at Austin. T.C. is supported by the U.S. DOE, Office of Basic Energy Sciences, under Contract No. DE-SC0025327. An award of computer time was provided by the INCITE program. This research used resources of both the Argonne and Oak Ridge Leadership Computing Facilities, which are DOE Office of Science User Facilities supported under contracts DE-AC02-06CH11357 and DE-AC05-00OR22725. Computational resources were also provided by National Energy Research Scientific Computing Center, which is a DOE SC User Facilities supported under contract DE- AC0205CH11231, and by Texas Advanced Computing Center, which is supported by U.S. National Science Foundation (NSF) under Grant No. OAC-1818253. Z.L. thanks Chen Hu for helpful discussions.

**Author contributions**

Z.L. conceived and directed the research. J.Y.Y., C.E.H., Z.Z., B.Z., M.D.B., and Z.L. developed the advanced computational functionalities and workflow. J.Y.Y. and C.E.H. performed the calculations. J.Y.Y., C.E.H., S.G.L., and Z.L. analyzed the data. All authors discussed the results and contributed to the writing of the manuscript.

**Competing interests**

The authors declare no competing interests.

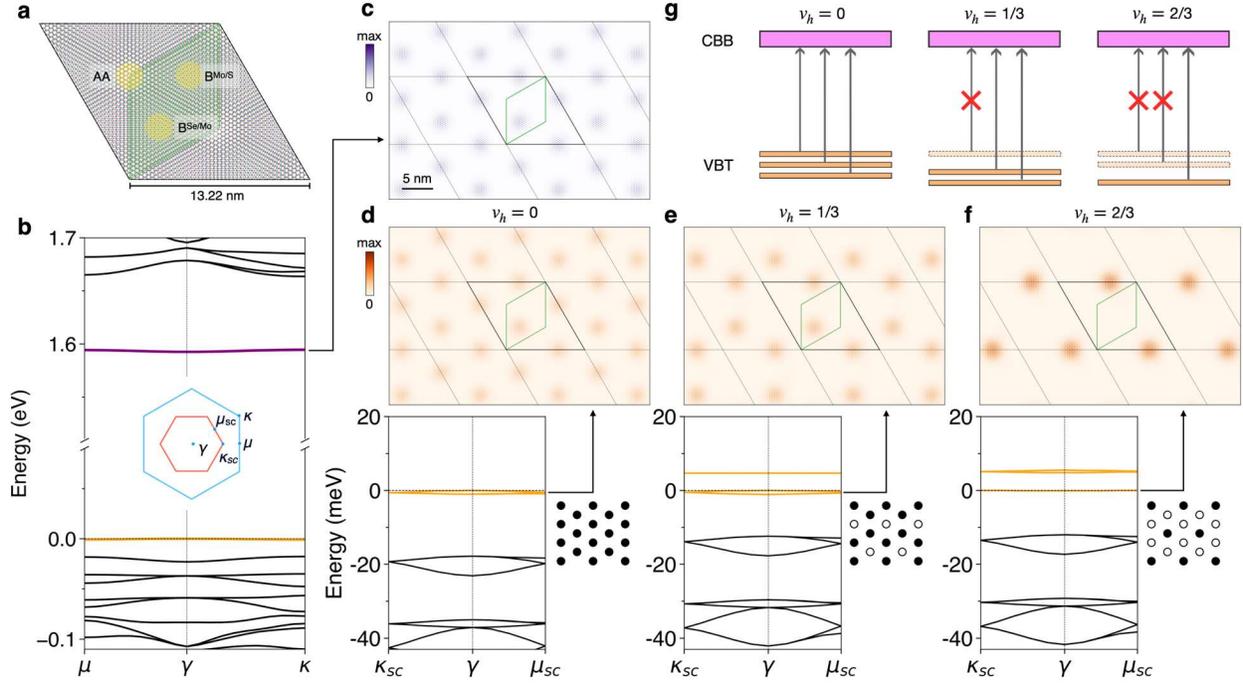

**Fig. 1. Single-particle electronic structure of generalized Wigner crystals. a.** Crystal structure of the angle-aligned MoSe$_2$/MoS$_2$ moiré superlattice. The green parallelogram represents the moiré unit cell, and the black parallelogram is a $\sqrt{3} \times \sqrt{3}$ supercell of the moiré unit cell. **b.** Band structure of the undoped ($v_h$ = 0) moiré superlattice plotted in the Brillouin zone of the moiré unit cell (blue hexagon). VBT complex is highlighted in orange, and CBB complex is highlighted in purple. **c.** Density map $\rho_{CBB}(\mathbf{r})$ of wavefunctions of the states in CBB. **d.** Density map $\rho_{VBT}^{occ}(\mathbf{r})$ of the wavefunctions of the occupied states in VBT, along with the band structure shown in the Brillouin zone of $\sqrt{3} \times \sqrt{3}$ moiré supercell (orange hexagon in **b**) for $v_h$ = 0. **e** and **f**. Similar to **d**, but for $v_h$ = 1/3 and $v_h$ = 2/3 hole doping concentrations, respectively. Symmetry-breaking Wigner-crystal ground states arise upon hole doping within VBT. **g.** Schematics of interband optical transitions between VBT and CBB at different hole doping concentrations.

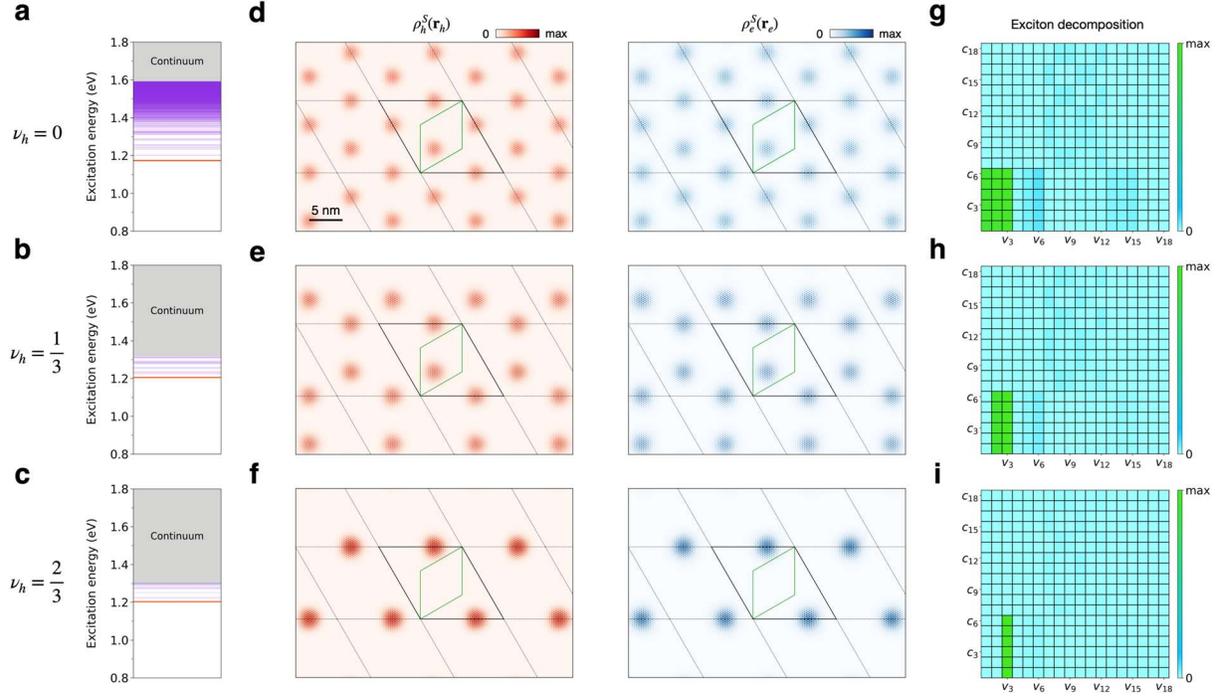

**Fig. 2. Excitonic internal structure of WCEs. a–c.** Excitation energies (color lines) of bound excitons at different hole doping concentrations, where the lowest-energy exciton states are indicated by red lines and higher energy states by purple lines. Continuum of free electron-hole pairs is indicated by the grey area, with its bottom representing the quasiparticle band gap. **d–f.** Excited hole density $\rho_h^S(\mathbf{r}_h)$ and excited electron density $\rho_e^S(\mathbf{r}_e)$ of the lowest-energy excitons at doping concentrations $v_h$ = 0, 1/3 and 2/3, respectively. WCEs arise upon fractional doping at $v_h$ = 1/3 and 2/3. **g–i.** Exciton decomposition $P_{vc}^S$ of the lowest-energy excitons in the basis of electron-hole (c-v) pairs at doping concentrations $v_h$ = 0, 1/3 and 2/3, respectively. The indexing of conduction and valence bands is fixed to the undoped case for ease of discussions. The excitons are mostly composed of electron-hole pairs from the CBB ($c_1 - c_6$) and VBT ($v_1 - v_3$) complexes.

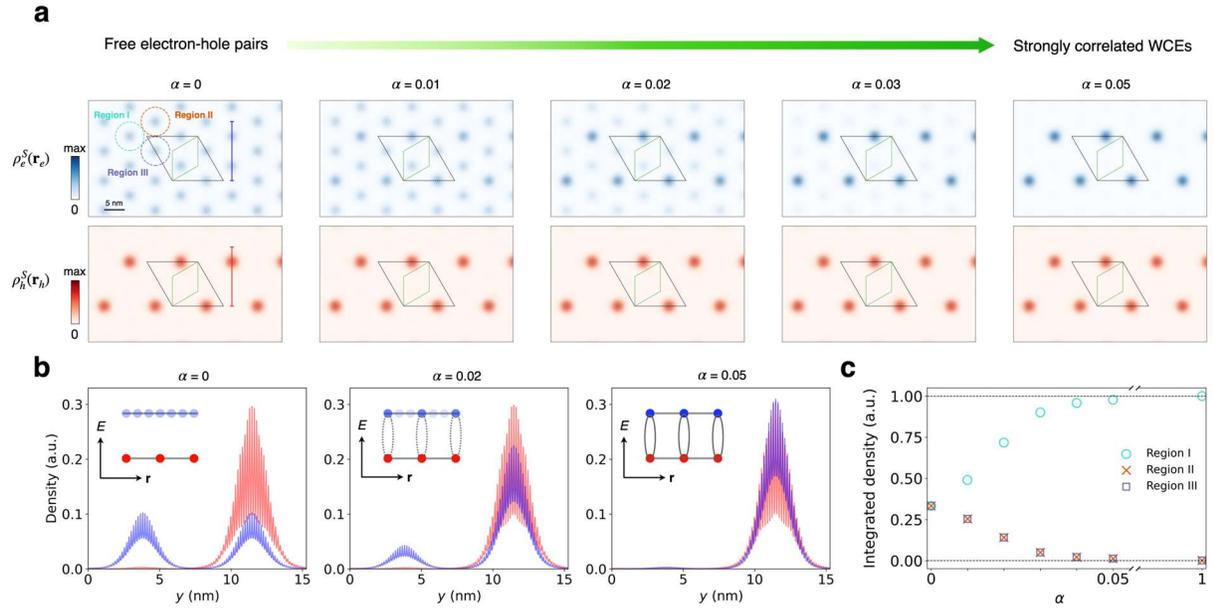

**Fig. 3. Electron-hole correlations of WCEs. a.** Computed excited hole density $\rho_h^S(\mathbf{r}_h)$ and excited electron density $\rho_e^S(\mathbf{r}_e)$ of the lowest-energy WCEs as a function of scaled electron-hole interaction kernel by the factor $\alpha$, for $v_h = 2/3$. At $\alpha = 0$, the electron-hole interaction is completely turned off, and the system is kinetic-energy driven and behaves as uncorrelated free electron-hole pairs. As $\alpha$ increases, electron-hole interaction is gradually turned on, leading to the formation of strongly bound and highly electron-hole correlated WCEs. $\alpha = 1.0$ gives the full electron-hole interaction strength of the system. **b.** Line cuts of the hole density (red) and electron density (blue) along the path indicated in the left-most panels of **a**. Insets show schematics of the development of increasing electron-hole correlations. **c.** Integrated density within the specific $B^{Se/Mo}$ stacking regions indicated in the top-left panel of **a**, as a function of $\alpha$. There are three such regions per supercell (i.e., per Wigner-crystal unit cell), labeled as I, II, and III in **a**. Region I is where the excited hole resides in the Wigner crystal phase. The sum of the depletion in the integrated electron density in regions II and III accounts for the density increase in region I as a function of increasing $\alpha$.

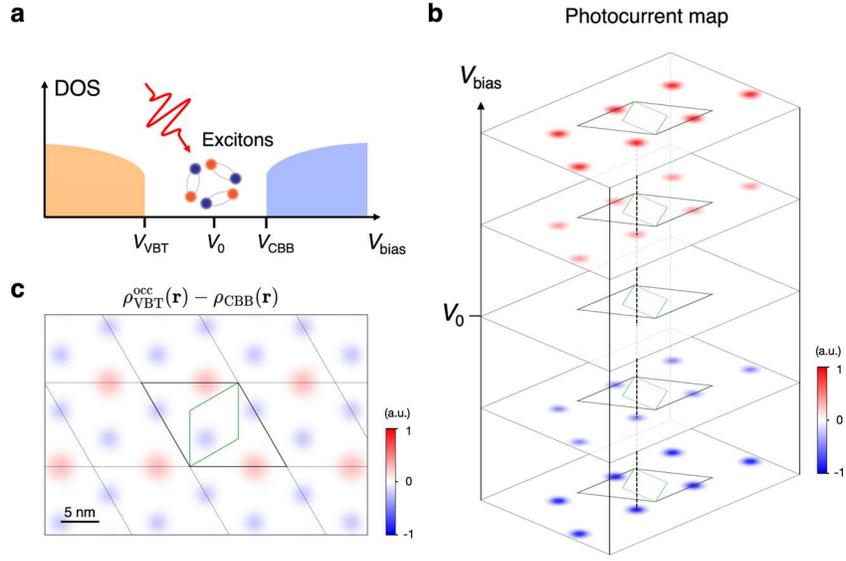

**Fig. 4. Proposed experimental demonstration of the internal structure and strongly correlated nature of WCEs. a.** Schematics of single-particle density of states (DOS) of a multi-band moiré superlattice system in 2D. In the PTM experiment[38] with laser on, electron and hole currents associated with the created excitons can be drawn with bias voltage ($V_{bias}$) in the gap region. $V_0$ represents a tip bias where the integrated electron and hole currents cancel out. **b.** Simulated photocurrent map as a function of bias voltage around $V_0$ for $v_h$ = 2/3 with laser on. The signed intensity map is taken as $I(\mathbf{r}) = \gamma \rho_h^S(\mathbf{r}) - (1-\gamma)\rho_e^S(\mathbf{r})$ with $\gamma$ = 1.0, 0.75, 0.5, 0.25, 0.0 from top to bottom, respectively, for the lowest-energy WCEs from the main gap. $I(\mathbf{r})$ is a simplified illustration of the experimental photocurrent[38] (neglecting factors such as tip configurations and tunneling matrix element effects). The vertical dashed line highlights the onsite sign change of the photocurrent with varying $V_{bias}$, as a direct signature of the strong excitonic correlation of WCEs. **c.** Uncorrelated free electron-hole pair density map $I'(\mathbf{r}) = \rho_{VBT}^{occ}(\mathbf{r}) - \rho_{CBB}(\mathbf{r})$, which may be compared with experimental data of scanning tunneling maps of VBT and CBB measured with laser off.